\documentclass[conference]{IEEEtran}
\IEEEoverridecommandlockouts
\usepackage{cite}
\usepackage{amsmath,amssymb,amsfonts}
\usepackage{algorithmic}
\usepackage{graphicx}
\usepackage{float}
\usepackage{subcaption}
\usepackage{amsmath}
\usepackage{textcomp}
\usepackage{xcolor}
\usepackage{subcaption}
\usepackage{caption}
\usepackage{multirow}
\def\BibTeX{{\rm B\kern-.05em{\sc i\kern-.025em b}\kern-.08em
    T\kern-.1667em\lower.7ex\hbox{E}\kern-.125emX}}

\begin{document}


\title{Efficient Feature Extraction and Classification Architecture for MRI-Based Brain Tumor Detection and Localization}





\author{

\IEEEauthorblockN{
    Plabon Paul\textsuperscript{\rm 1},
    Md. Nazmul Islam\textsuperscript{\rm 2},
    Fazle Rafsani\textsuperscript{\rm 2,3},
    Pegah Khorasani\textsuperscript{\rm 4},
    Shovito Barua Soumma\textsuperscript{\rm 2,4}
}
\IEEEauthorblockA{
    Department of Mechanical Engineering\textsuperscript{\rm 1},\\
    Department of Computer Science and Engineering\textsuperscript{\rm 2},\\ 
    Bangladesh University of Engineering \& Technology, Dhaka, Bangladesh\\
    School of Computing and Augmented Intelligence\textsuperscript{\rm 3}, Arizona State University\\
    College of Health Solutions\textsuperscript{\rm 4}, Arizona State University\\
    Email: plabonpaul.97@gmail.com, 1605088@ugrad.cse.buet.ac.bd, \{frafsani, pkhorasa, shovito\}@asu.edu
    }
}

\maketitle

\begin{abstract}
Uncontrolled cell division in the brain is what gives rise to brain tumors. If the tumor size increases by more than half, there is little hope for the patient's recovery. This emphasizes the need for rapid and precise brain tumor diagnosis. MRI imaging plays a crucial role in analyzing, diagnosing, and planning therapy for brain tumors. A brain tumor's development history is crucial information for doctors to have. When it comes to distinguishing between human soft tissues, MRI scans are superior. Deep learning is one of the most practical methods for quickly getting reliable classification results from MRI scans. Early human illness diagnosis has been demonstrated to be more accurate when deep learning methods are used. In the case of diagnosing a brain tumor, when even a little misdiagnosis might have serious consequences, accuracy is especially important.  Disclosure of brain tumors in medical images is still a difficult task. Also because of the scarcity of annotated data localizing the tumor regions is quite expensive task. Brain MRIs are notoriously imprecise in revealing the presence or absence of tumors. Using MRI scans of the brain, a Convolutional Neural Network (CNN) was trained to identify the presence of a tumor in this research. Results from the CNN model showed an accuracy of 99.17\%. The CNN model's characteristics were also retrieved and we also localized the tumor regions from the unannotated images using GradCAM, a deep learning explainability tool. In order to evaluate the CNN model's capability for processing images, we applied the features via the following machine learning models: KNN, Logistic regression, SVM, Random Forest, Naive Bayes, and Perception. CNN and machine learning models were also evaluated using the standard metrics of Precision, Recall, Specificity, and F1 score. The significance of the doctor's diagnosis enhanced the accuracy of the CNN model's assistance in identifying the existence of tumor and treating the patient.
\end{abstract}


\section{Introduction}
Brain is a core part of CNS (Central Nervous System), which regulates all physiological and cognitive activities i.e. thought, emotion, touch, motor skils, vision, respiration etc.~\cite{8664160}.
Brain tumor is uncontrolled cell division in brain or other parts of the CNS that causes malfunctioning. The malignancy of a tumor depends on how fast the cell reproduces. Non-malignant (benign and not cancerous) tumors grow slowly and do not spread into other tissues. Malignant brain tumors are cancerous. Most of the time, they multiply and invade neighboring healthy tissues~\cite{brain-edu}~\cite{Jeffrey}. In USA, there are an estimated 20,500 primary brain tumors identified each year; 3750 of these instances include people under the age of 19 and 2870 involved children under the age of 15~\cite{Hoskinson}. In comparison to tumors in any other organ of the human body, diagnosing a tumor in the brain is particularly difficult. The blood-brain barrier (BBB) surrounds the brain, making it impossible for regular radioactive markers to detect the tumor cells' increased activity. Additionally, tumor size, shape, location, and type make early detection more challenging~\cite{Jeffrey,Aboody}. 

Deep learning and machine learning are two new technologies that have significantly developed different fields of applications~\cite{soumma2024SSL,dip2024,10054906, soumma2024wearablebasedrealtimefreezinggait}. 
An important aspect of this study area involves automating the segmentation and categorization of brain tumors. 
Akram et. al~\cite{akram2011computer} introduced a computer-aided method for identifying tumors in which they segmented tumors using the global thresholding methodology.


Parasuraman et al\cite{kumar2019brain} used a feed-forward neural network to classify tumor and normal regions. The method involves four steps: pre-processing with filters, image segmentation with clustering, feature extraction using gray-level co-occurrence matrix (GLCM), and tumor classification with ensemble classifier. 
Irmak et al.~\cite{irmak2021multi} developed three CNN models for three different datasets. 
However, training such models and validation requires a significant amount of pixel level annotated data and in the field of medical imaging getting such data is too costly and requires domain experts. 

\begin{figure*}[h]
\centering
\includegraphics[width=0.83\textwidth]{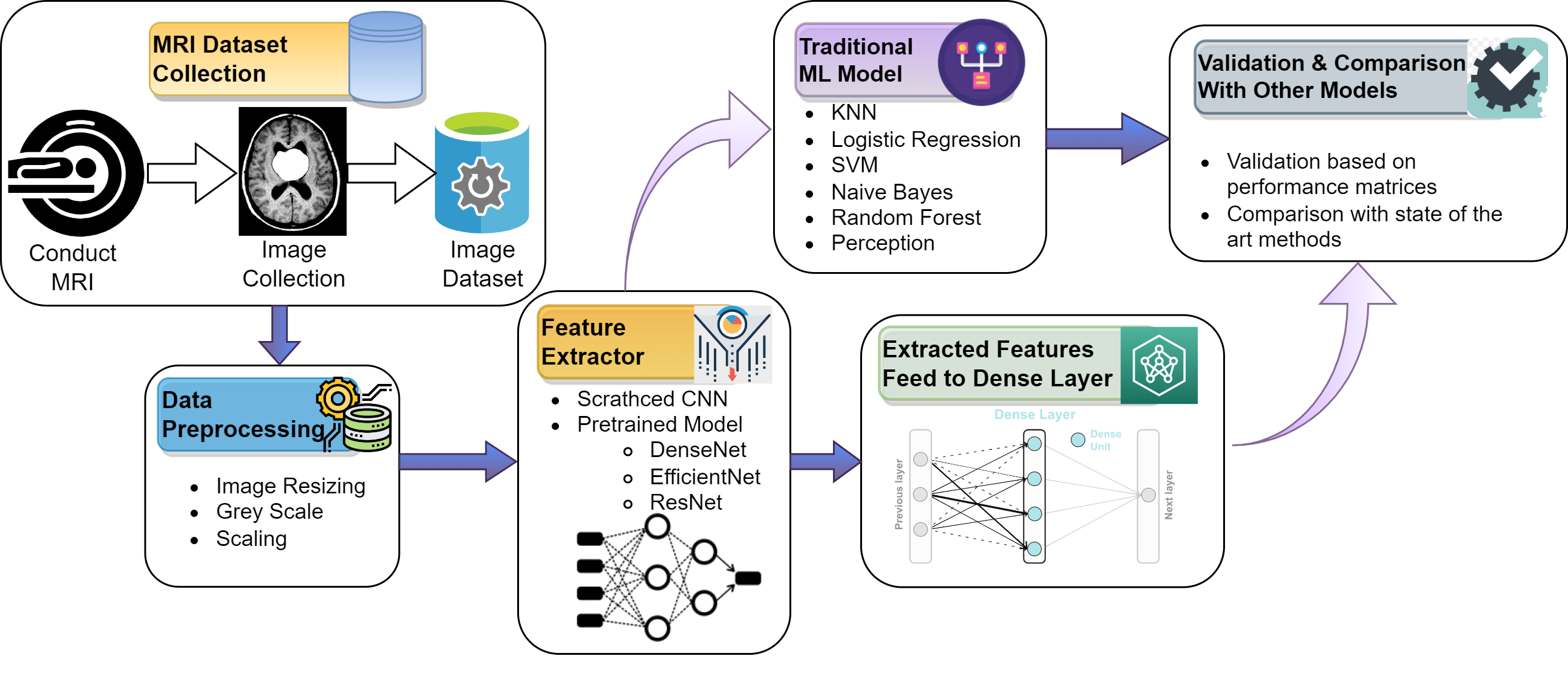}
\caption{Workflow of the suggested deep learning approach demonstrating many essential phases, such as preprocessing, feature extraction, classification, and model validation for brain tumor }
\label{fig:method}
\end{figure*}

In this study, we proposed a convolution Neural Network based architecture that aims to detect tumorous brain MRI images and localize the affected regions. We achieved an accuracy of 99.83\% without pixel level annotated data. Finally we applied the technique proposed in GradCAM \cite{selvaraju2020grad} for producing visual explanations for decisions from our CNN model. This method localizes the tumorous region from the gradients of the final convolution layer of the model which eliminates the need for pixel level annotation for localizing the anomalous regions.

\section{Materials and Methods}
We implemented a CNN model from scratch, three pre-trained models, and five traditional ML models in our proposed method. Our main objective is to diagnose brain tumors effectively and precisely by sending MRI pictures of the tumors to a CNN and localize the anomalous regions. Fig.~\ref{fig:method} represents the workflow of our study. Labeled MRI images are supplied into a CNN feature extractor after minimum preprocessing, and the retrieved features are fed into a classification layer. Finally, we compared the performance of our trained model to the current state of the art in various aspects.

\subsection{Dataset}
\begin{figure}[!t]
\centering
\includegraphics[scale=0.3]{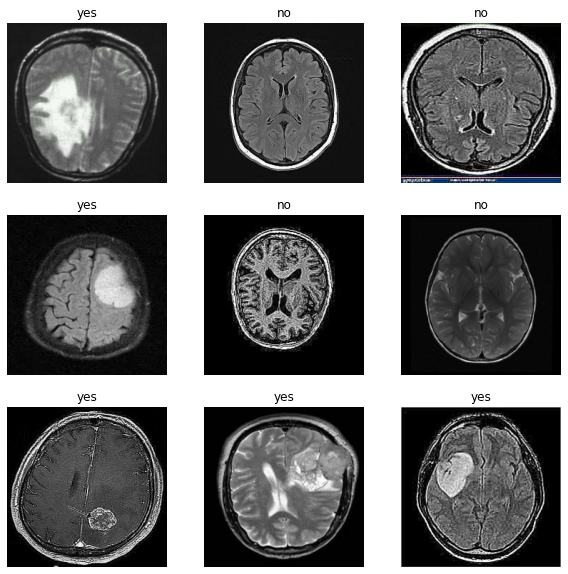}
\caption{Samples of the brain MRI dataset}
\label{fig:samples}
\end{figure}

The Brain Tumor Detection 2020 (BR35H)~\cite{Br35H} dataset, which includes two unique classes of MRIs of brain tumors (1500 negative and 1500 positive), is utilized to train CNN.80\% of the images from this dataset are used for training the model. Fig \ref{fig:samples}  displays a few samples from datasets that includes data from two different types of brain MRI. 
 
\subsection{Data Preprocessing}
All the images are preprocessed before being fed to CNN, as described in Fig.~\ref{fig:method}  These images are initially transformed into single-channel images, sometimes referred to as greyscale images. The dimension of each image is different in the dataset. Therefore, these images are reshaped into a fixed size. Then, every image is converted to a two-dimensional array. After that, each image is normalized so that the value of each array element is converted to the range [0,1].

\subsection{Feature Extraction}
\subsubsection{Modified DenseNet}
DenseNet is a type of network architecture in which each layer is directly connected to every other layer (within each dense block)\cite{huang2017densely}. For each layer, the feature maps of all preceding layers are treated as separate inputs, whereas its own feature maps are passed on as inputs to all subsequent layers. The original network's inputs are 256*256 in size. This model has 707 layers and it is fine tuned after 500 layer for better performance. A global average pooling layer and a dense layer have also been added, which enhance the performance of the original model.

\subsubsection{Modified ResNet50}
The Residual Networks, or ResNet for short, took first place in the 2015 ImageNet competition\cite{he2016residual} and is now employed for a variety of computer vision-related applications. Here, solving the issue of vanishing gradients and drastically reducing the number of parameters is the major objective of training a very deep neural network. As shown in Fig.~\ref{fig:ResNet50},The layer connections are skipped. The size of the inputs in the original architecture was 224*224. It requires 175 layers  where after 125 layer rest of them are fine tuned. A global average pooling layer and a dense layer have been added, which enhance the performance of the original model.

\subsubsection{Modified EfficientNetB0}
Tan et al. developed the EfficientNetB0 architecture to make scaling models simpler by balancing the network's height, depth, and input resolution to increase accuracy\cite{tan2019efficientnet}. This study uses the underlying EfficientNet-B0 network, which is based on the MobileNet-V2 inverted bottleneck residual blocks. The size of the inputs in the original network is 224*224 and here in the modified network it has been changed to 64*64. Having 237 layers in this architecture, rest of the layer after 150th layer were fine tuned. In addition, a dense layer and a global average pooling layer have been added to the original model to improve its performance.

\subsubsection{CNN model from Scratch}
\begin{figure}[!t] 
\centering
   \begin{subfigure}[b]{0.54\textwidth}
   \includegraphics[width=0.9\textwidth]{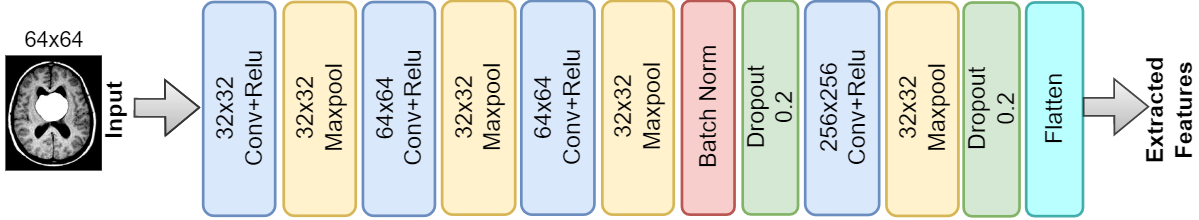}
  \caption{}
   \label{fig:cnn} 
\end{subfigure}
\begin{subfigure}[b]{0.54\textwidth}
   \includegraphics[width=0.9\textwidth]{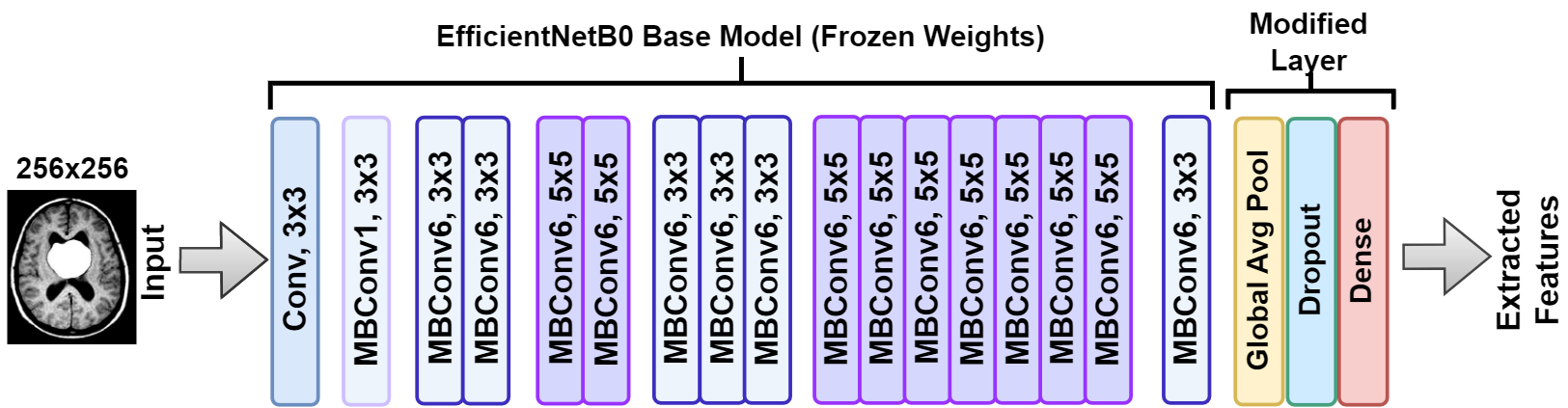}
   \caption{}
   \label{fig:EfficientNetB0}
\end{subfigure}
\begin{subfigure}[b]{0.54\textwidth}
   \includegraphics[width=0.9\textwidth]{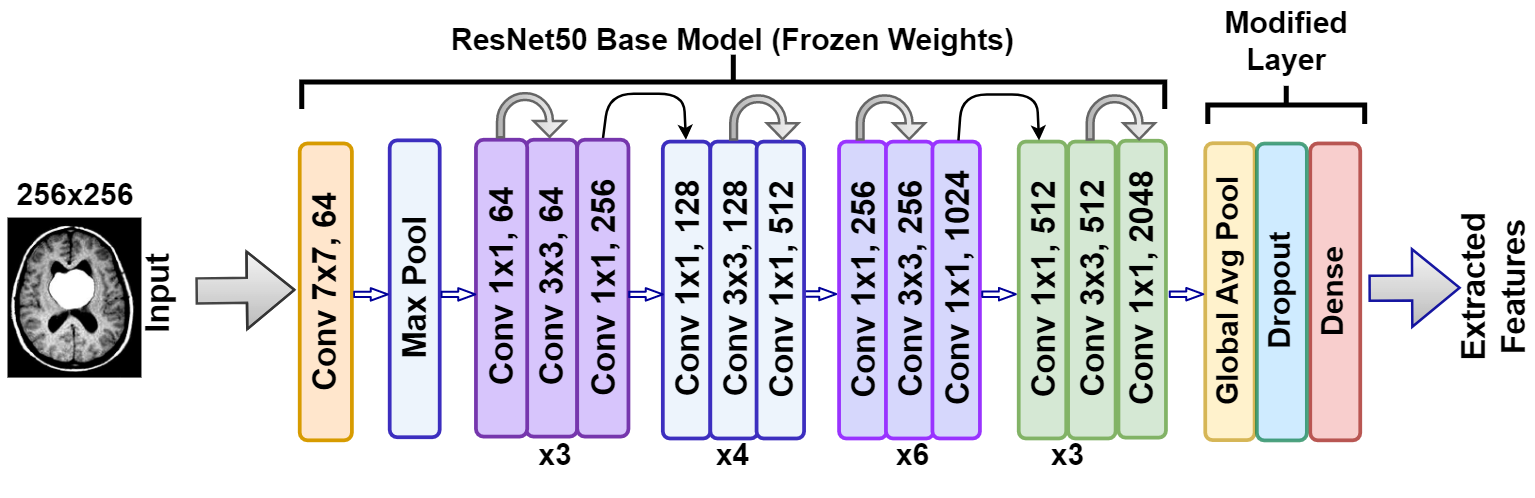}
  \caption{}
   \label{fig:ResNet50}
\end{subfigure}
\begin{subfigure}[b]{0.54\textwidth}
   \includegraphics[width=0.9\textwidth]{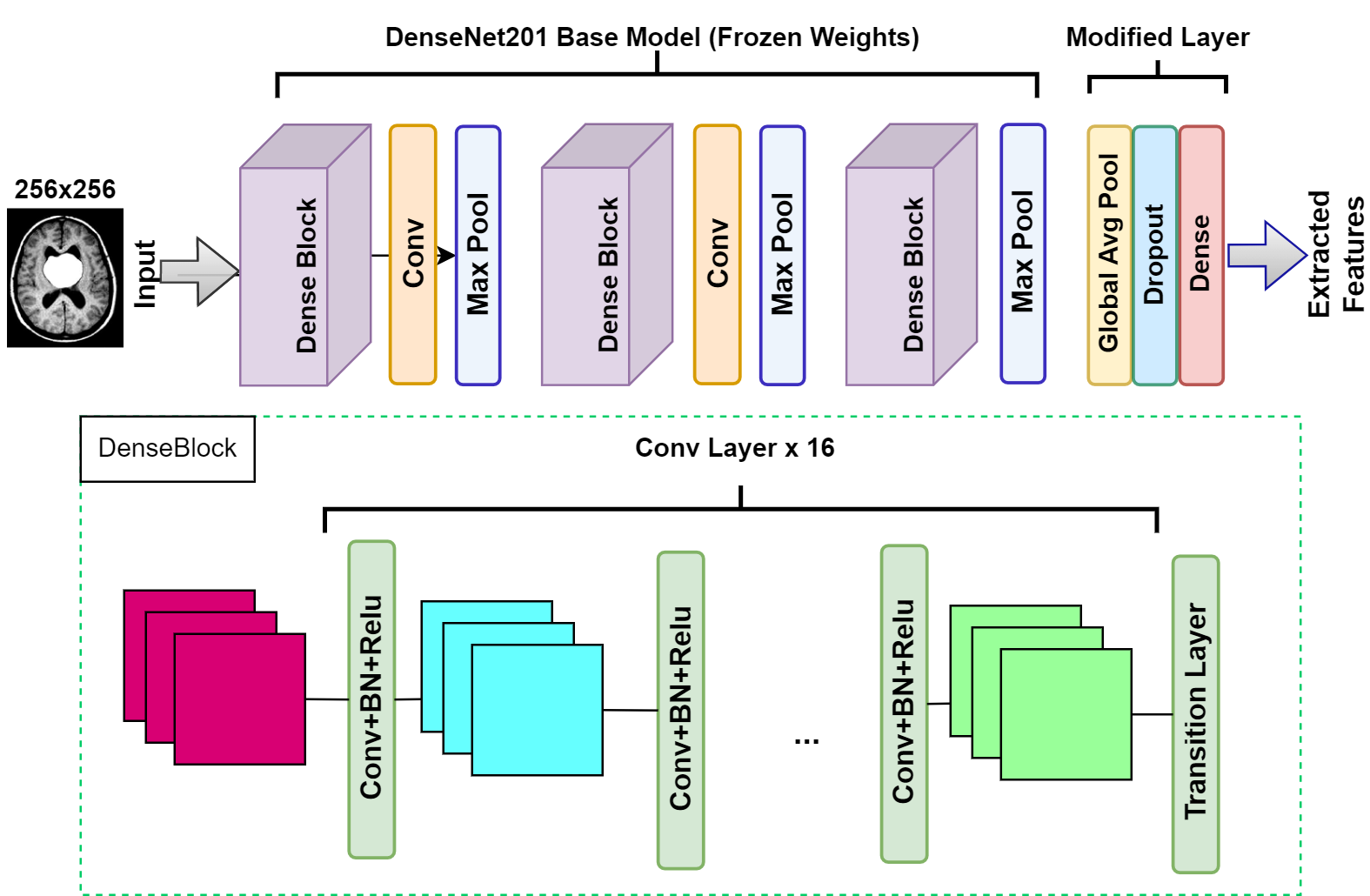}
  \caption{}
   \label{fig:DenseNet201}
\end{subfigure}

\caption{CNN feature extractors used in this paper (a) Proposed 12 layers scratch-built CNN model. (b) Modified EfficientNetB0. (c) Modified ResNet50 (d) Modified DenseNet201}

\end{figure}

Fig.~\ref{fig:cnn} shows the architecture of a twelve layer CNN model that is built from scratch to classify brain tumors. This model was built by trying different ways to tune the hyper-parameters. The best one is selected by running a random search on various combinations of the model's hyper-parameters while using Keras Tuner. The model has four convolution layers, four max pooling layers, one batch normalization layer, one flatten layer, and one dropout layer. Here the input size is 32*32. After passing through all the convolution, max pooling, batch normalization, and dropout layers, the input size finally becomes width*height. In every convolution layer, the RELU activation function was utilized. And for optimization, the Adam optimizer was used with the Sparse Categorical cross-entropy loss function.

\subsection{Machine Learning Classifiers}
\subsubsection{KNN} 
KNN is a Supervised learning algorithm which can solve both classification and regression-predicting problems. It is a lazy algorithm because it uses all of the data for training during classification and lacks a training phase. 

\subsubsection{SVM}
SVM focuses solely on the training samples that are situated closest to the ideal class border in the feature space.
The name of the approach comes from these samples, which are known as support vectors. The SVM classifier is fundamentally binary in that it recognizes only one distinction between two classes. 

\subsubsection{Naive Bayes}
Naive Bayesian networks (NB) are made up of DAGs with a single parent (representing the invisible node) and a large number of children (corresponding to the visible nodes), and they make a strong assumption that each child node is independent of the other child nodes. As a result, the independence model is based on estimating:

\begin{equation} \label{eq1}
\begin{split}
    R = \frac{P(i|X)}{P(j|X)}=\frac{P(i)P(X|i)}{P(j)P(X|j)}= \frac{P((i) \Pi P(X_r|i) }{P((j) \Pi P(X_r|j)}
\end{split}
\end{equation}

If R$>$1, then the Naive Bayes model will predict \textit{i}, else it will predict \textit{j}. Typically, Bayes classifiers perform less accurately than other, more advanced learning algorithms (such as ANNs). 
\subsubsection{Random Forest}
Random Forest is a classifier that employs several decision trees on various subsets of the input dataset and averages the outcomes to improve the projected accuracy of the dataset.
The final class is determined by majority voting across all of the trees. 

\subsubsection{Multi Layer Perceptron}
This classifier solves a quadratic programming issue with linear constraints as opposed to the non-convex, unconstrained minimization problem that neural networks often use to find their weights. 
In a multilayer perceptron, neurons are structured in layers and totally coupled with one another by edges to create a directed graph.
The input layer, a number of hidden layers, and the output layer are the layers of multilayer perceptron neural networks. There is a summation function and an activation function for every neuron.


\section{Results and Discussion}
\begin{figure}
    \centering
    \subfloat[ResNet50]{{\includegraphics[width=4.2cm]{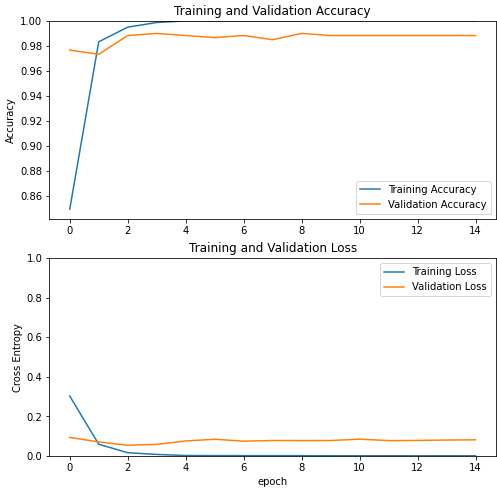} }}
    \subfloat[EfficientNetB0]{{\includegraphics[width=4.2cm]{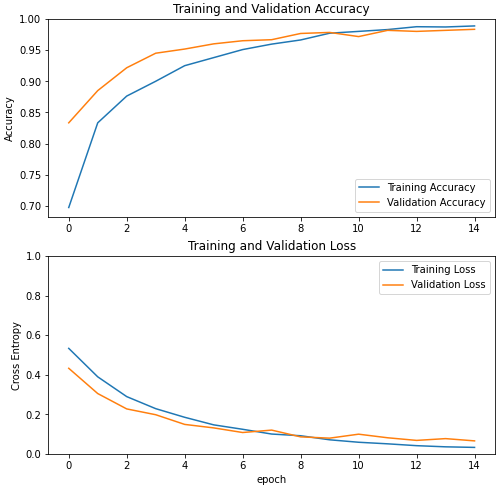} }}
    \\
    \subfloat[DenseNet]{{\includegraphics[width=4.2cm]{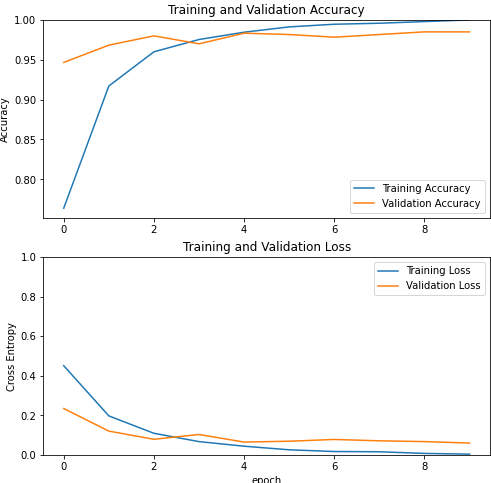} }}
    \subfloat[Scratch CNN]{{\includegraphics[width=4.2cm]{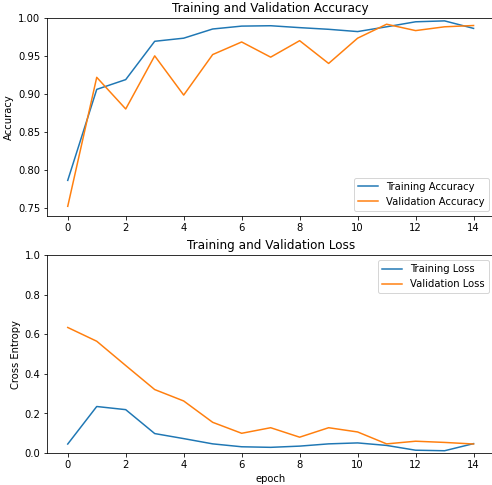} }}
    \caption{Accuracy and Loss curve for different models}
    \label{fig:Accuracy-Loss}
\end{figure}
\begin{figure}[!t] 
\centering
   \begin{subfigure}[b]{0.55\textwidth}
   \includegraphics[width=0.85\textwidth]{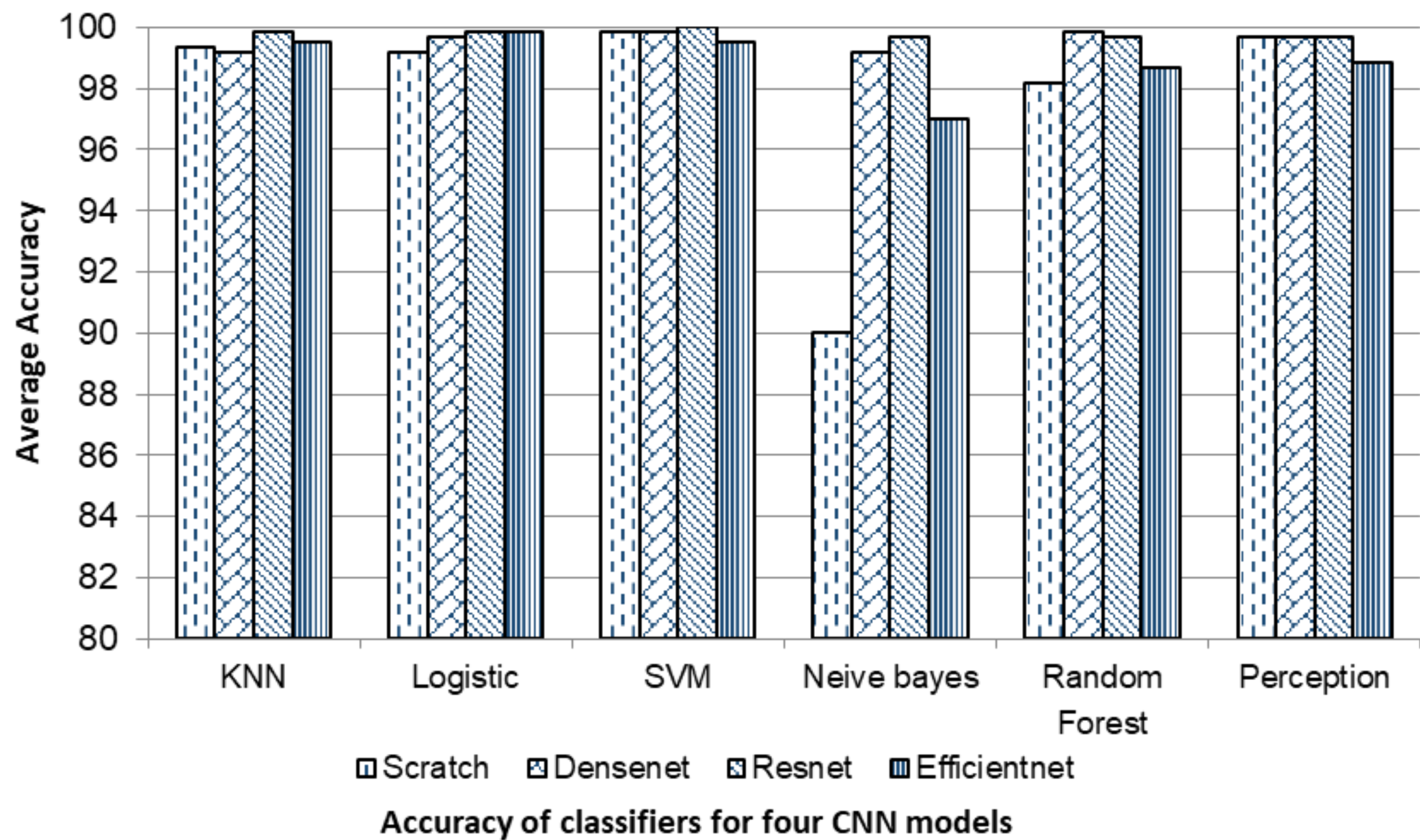}
   \caption{Accuracy of Classifiers for CNN Models}
   \label{fig:Accuracy-classifiers} 
\end{subfigure}
\begin{subfigure}[b]{0.55\textwidth}
   \includegraphics[width=0.85\textwidth]{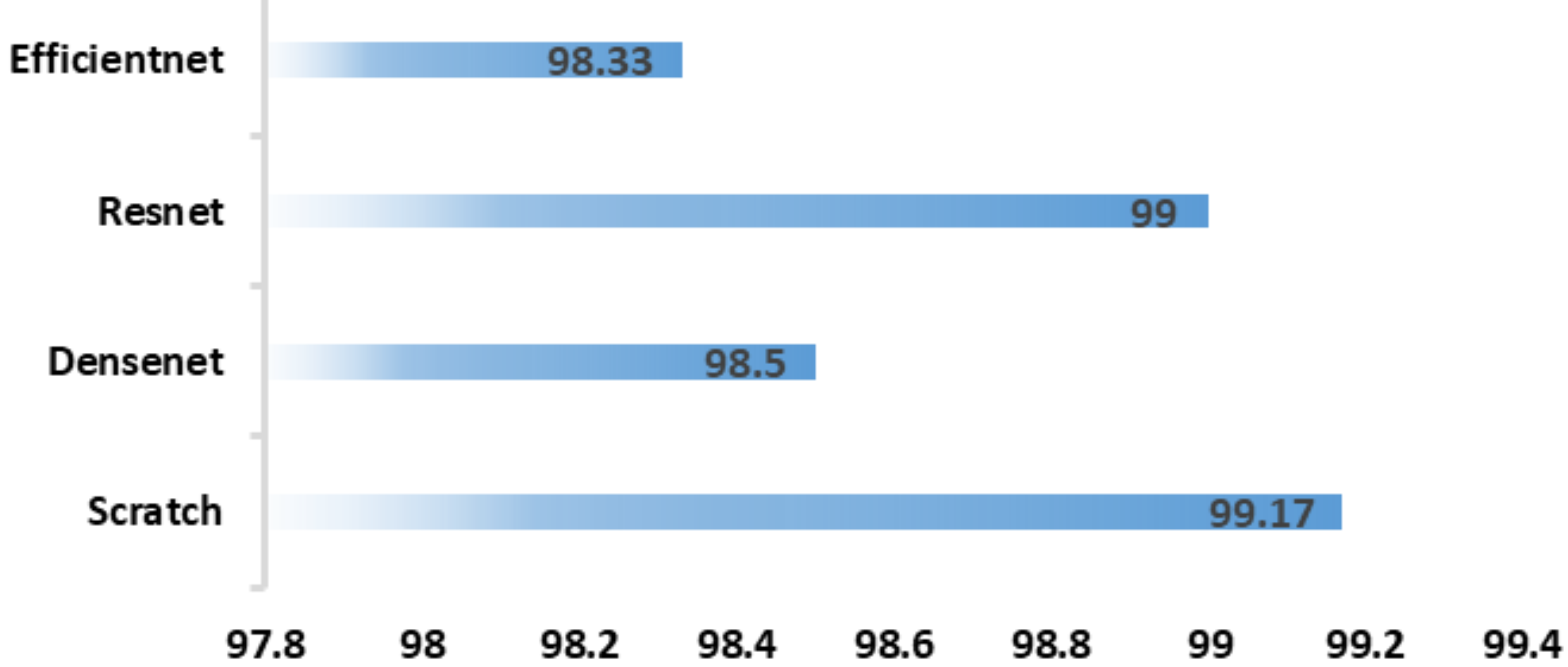}
   \caption{Accuracy of CNN models}
   \label{fig:Accuracy-CNN}
\end{subfigure}
\caption{Classification accuracy after feature extraction}
\end{figure}

In this research study, we have implemented four different models which are Densenet, Resnet50, EfficientnetB0, and our own CNN model. For optimization of our deep learning model, we used Adam algorithm. First, the performance is evaluated using several performance metrics. During the training of a model, we concentrated on reducing loss while simultaneously boosting accuracy. Table~\ref{tab:Accuracy-CNN} displays the validation accuracy of each model.  Here, we can observe that the validation accuracy was best achieved by our own CNN model. Fig.~\ref{fig:Accuracy-CNN} also displays the outcome graphically. Then, using these models, extracted features are sent to various machine learning models. And the outcome is revealed on Fig~\ref{fig:Accuracy-classifiers}. From the Table~\ref{tab:Accuracy-Models}, we can conclude that Scratch+SVM, DenseNet+SVM, ResNet+SVM, and EfficientNet+Logistic Regression perform better than any other combination of machine learning and pretrained models and the comparison is illustrated in Fig.~\ref{fig:Accuracy-classifiers}.
\begin{figure*}[h]
\centering
\includegraphics[width=0.8\textwidth]{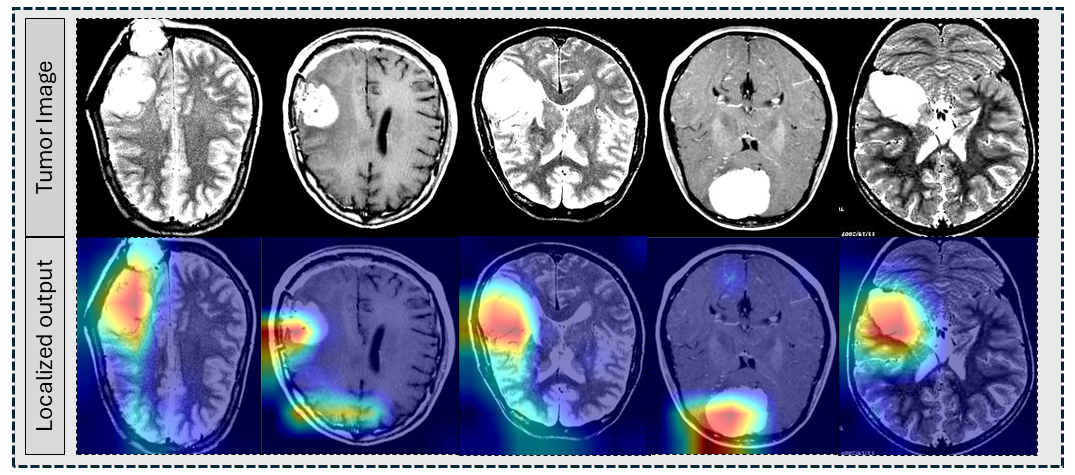}
\caption{Localization of tumorous region in Brain MRI}
\label{fig:localization}
\end{figure*}
\subsection{Performance metrics}
Evaluation metrics are used to measure the quality of a model. A model can be tested using a wide variety of evaluation metrics. Confusion metrics is an N*N matrix, where N is the number of expected classes, is a confusion matrix. Since N=2 applies to our issue, we receive a 2*2 matrix. True Positive (TP) indicates the number of accurately categorized attack records. True Negative (TN) is the proportion of correctly categorized normal records. False Positive (FP) is the quantity of typical records that were misclassified. False negatives (FN) are the number of attack records that were wrongly categorized.
\begin{align} 
&\mathit{Accuracy} = \frac{TP+TN}{TP+TN+FP+FN}\\
&\mathit{Precision} = \frac{TP}{TP+FP}\\
&\mathit{Recall} = \frac{TP}{TP+FN}\\
&F1 = \frac{2*\mathit{Precision}*\mathit{Recall}}{\mathit{Precision}+\mathit{Recall}}\\
&\mathit{Specifity} = \frac{TN}{TP+FP}
\end{align}
\begin{figure}[!h]
  \centering
  \begin{minipage}[b]{0.55\textwidth}
    \includegraphics[width=0.9\textwidth]{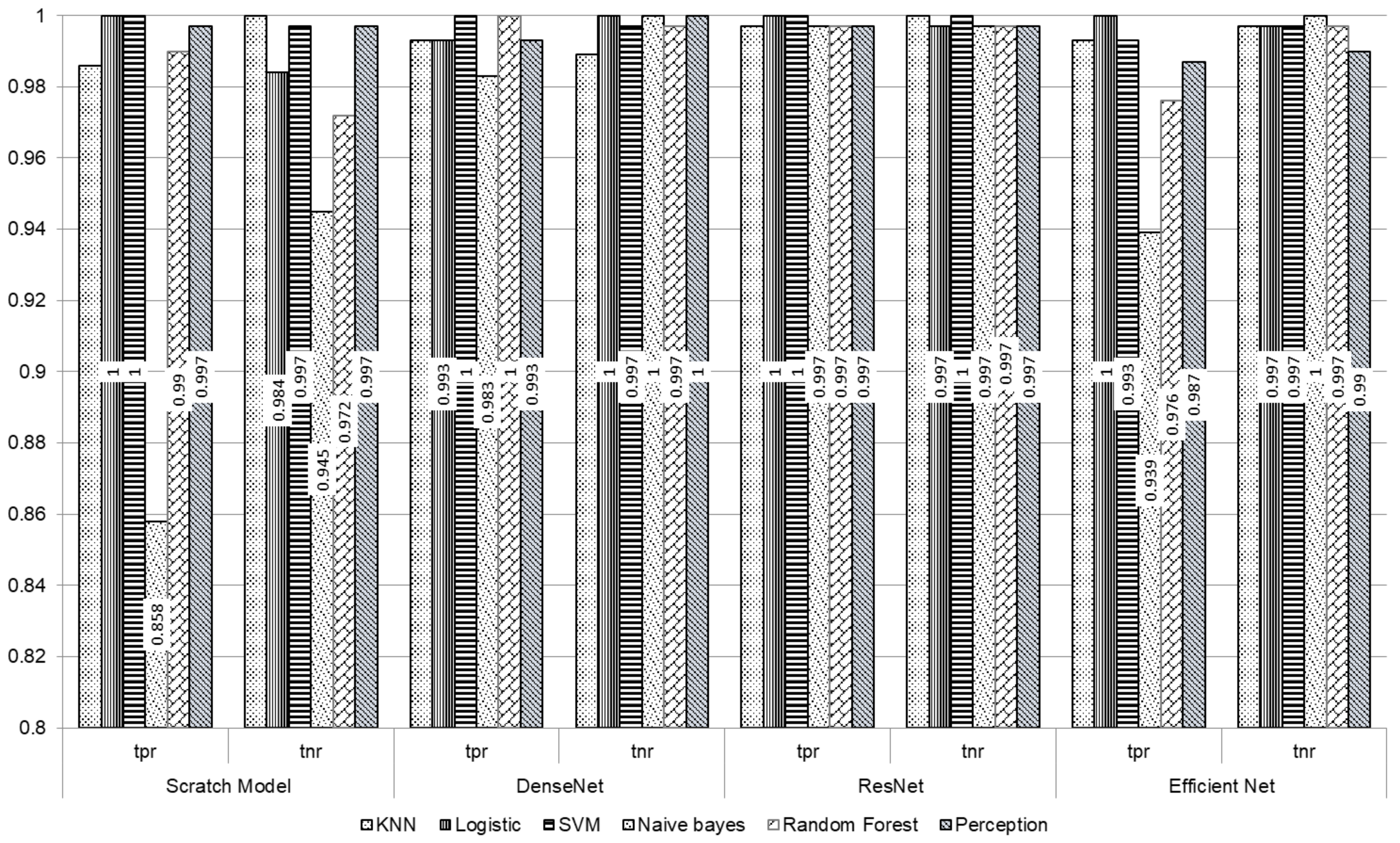}
    \caption{TPR and TNR of each model.}
    \label{fig:classifiers_tpr}
  \end{minipage}
  \begin{minipage}[b]{0.55\textwidth}
    \includegraphics[width=0.9\textwidth]{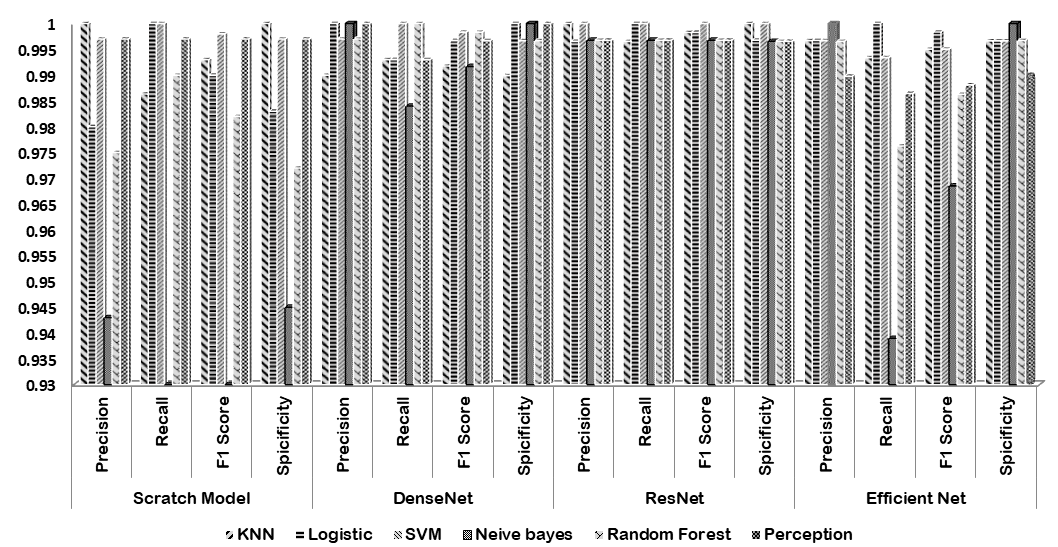}
    \caption{Classification metrics of each model.}
    \label{fig:classifiers_matrics}
  \end{minipage}
\end{figure}

\begin{table}[b]
\caption{Accuracy of classifiers for CNN models}
\begin{center}
\resizebox{\columnwidth}{!}{%
    \begin{tabular}{|c|c|c|c|c|}
 \hline
& Scratch CNN & DenseNet & ResNet50 & EfficientNetB0	 \\ 
 \hline
 KNN &	99.33 & 99.167 & 99.83 & 99.5\\
 \hline
 Logistic & 99.17 & 99.67 & 99.83 & 99.83\\
 \hline
 SVM & 99.83 & 99.83 & 100 & 99.5\\
 \hline
 Naive Bayes & 90 & 99.17 & 99.67 & 97\\
 \hline
 Random Forest & 98.17 & 99.83 & 99.67 & 98.667\\
 \hline
 Perception & 99.67 & 99.67 & 99.67 & 98.83\\
 \hline
\end{tabular}
}
\end{center}
\label{tab:Accuracy-Models}
\end{table}
\subsubsection{Comparison with other works}
We will now compare our conclusions to a variety of other approaches that were suggested in the literature review. The comparative result is shown in Table~\ref{tab:Performance_comparison}. It covers the findings of some of the newest methods suggested for locating brain tumors. The table demonstrates that WCNN was used to achieve the highest accuracy, which was 99.3. The other methods are less accurate than this. Even if the accuracy of our own CNN model is less accurate than that of the Sarhan et al.\cite{sarhan2020brain}, it is still more accurate than alternative approaches. A number of pretrained models that we also trained showed impressive accuracy. Then, in order to increase the accuracy of our models, we mix these models with machine learning models. The highest accuracy we can achieve is 99.83\%, which is higher than the paper by Sarhan et al.\cite{sarhan2020brain}. 

\begin{table}[h]
    \caption{Performance comparison with other models}
        \begin{center}
        \resizebox{\columnwidth}{!}{%
        \begin{tabular}{|c|c|c|} 
         \hline
         Authors & Technique & Accuracy \\
         \hline
         Sarhan et al.\cite{sarhan2020brain} & WCNN &	99.3\\
        \hline
        Bhanothu et al.\cite{bhanothu2020detection} & Faster-R-CNN & 77.6\\
        \hline
        Ismael et aI.\cite{ismael2020enhanced} & ResNet50 & 99\\
        \hline
        Kaplan et aI.\cite{kaplan2020brain} & KNN & 95.56\\
        \hline
        Rehman et aI.\cite{rehman2020deep} & VGG16 & 98.69\\
        \hline
        Tahir et aI.\cite{tahir2019feature} & SVM & 86\\
        \hline
        Sethy et aI.\cite{sethy2021data} & VGG19+SVM & 97.89\\
        \hline
        Gajula et aI.\cite{gajula2021mri} & U-Net &	96.9\\
        \hline
        Ahmadia et aI.\cite{ahmadi2021detection} & CNN &	96\\
        \hline
        Saffar et aI.\cite{al2021hybrid} & MLP+SVM & 91.02\\
        \hline
        Kaldera et aI.\cite{kaldera2019brain} & Faster-R-CNN &	94\\
        \hline
        
        Ayadi et aI.\cite{ayadi2021deep} & CNN & 98.49\\
        \hline
        Sultan et aI.\cite{sultan2019multi} & DNN & 96.61\\
        \hline
        Kumar et aI.\cite{kumar2021multi} & Residual network and global average pooling	& 98.02\\
        \hline
        Abiwinanda et aI.\cite{abiwinanda2019brain} & CNN & 97.08\\
        \hline
        Deepak et aI.\cite{deepak2021automated} & CNN and SVM & 97.17\\
        \hline
        \textbf{Our Model} & \textbf{DenseNet and RF }& \textbf{99.83}\\
        \hline
        
        \end{tabular}
        }
        
        \end{center}
    \label{tab:Performance_comparison}
    \vspace{-4mm}
\end{table}

\subsection{Anomalous region localization}
We utilized the GradCAM method for localizing the tumorous regions of the images. The Gradient-weighted Class Activation Mapping (Grad-CAM) approach utilizes the gradients of a specified target class, which propagate into the final convolutional layer, to generate a coarse localization map that emphasizes significant areas in the image relevant to the class's prediction. Figure \ref{fig:localization} shows the model's robust output for the localization technique. The sample images are not annotated pixel by pixel. Only the labels are used to generate the localization heatmap for the images.

\section{Future Works and Conclusion}
In this study, we explored different models to detect brain tumor more effectively and precisely. We have surpassed all previously investigated methods with an accuracy of 99.83\%. We also proposed a tumor localization method which localizes tumorous regions without pixel level annotation. The robust localization of the regions and the accurate detection illuminates the path to use lightweight models for anomaly detection in medical images. Since brain tumor can lead to cancer, the impact of brain tumors are terrible for us and endanger our lives. We believe that our approach has the potential to reduce the risk of developing cancer and save many lives. In the future, we hope to develop a model that can accurately detect all types of tumors. Although our model solely analyzes MRI dataset, we also wish to take into account other medical imaging methods, such as CT (Computed Tomography) scan, PET (Positron-Emission Tomography) etc. 

\bibliographystyle{IEEEtran}
\bibliography{soumma}

\end{document}